\documentclass[aip,reprint,superscriptaddress,nofootinbib]{revtex4-2}

\usepackage[T1]{fontenc}
\usepackage[utf8]{inputenc}
\usepackage{amsmath,amssymb,amsthm,mathtools}
\usepackage{bm}
\usepackage{braket}
\usepackage{graphicx}
\usepackage{xcolor}
\usepackage{hyperref}

\newcommand{\Tr}{\operatorname{Tr}}
\newcommand{\I}{\mathbb{I}}
\newcommand{\id}{\operatorname{id}}
\newcommand{\C}{\mathbb{C}}
\newcommand{\R}{\mathbb{R}}

\newcommand{\Pos}{\mathrm{Pos}}
\newcommand{\CP}{\mathrm{CP}}
\newcommand{\KS}{\mathrm{KS}}

\newcommand{\diag}{\operatorname{diag}}

\newtheorem{theorem}{Theorem}
\newtheorem{lemma}{Lemma}

\theoremstyle{remark}
\newtheorem{remark}{Remark}

\begin{document}

\title{Sufficient conditions for the Kadison--Schwarz property of unital positive maps on $M_3$}

\author{Adam Rutkowski}
\affiliation{Institute of Theoretical Physics and Astrophysics \\ Faculty of Mathematics, Physics and Informatics, University of Gda\'{n}sk, 80-952 Gda\'{n}sk, Poland}

\date{\today}

\begin{abstract}
Kadison--Schwarz (KS) maps form a natural class of positive linear maps
lying strictly between positivity and complete positivity.
Despite their relevance in operator algebras and quantum dynamics,
explicit analytic sufficient conditions for the KS property remain scarce
beyond low-dimensional or highly symmetric settings.
In this work we analyze unital positive linear maps on $M_3$ within the
Bloch--Gell--Mann representation and derive explicit analytic
sufficient conditions ensuring the Kadison--Schwarz property.
The approach exploits unitary equivalence together with structural
properties of the Lie algebra $\mathfrak{su}(3)$ and does not rely on
numerical optimization or semidefinite-programming methods.
A key mechanism is the cancellation of contributions associated with
antisymmetric structure constants, which reduces the problem to
estimates governed solely by the symmetric tensor $d_{ijk}$.
The results clarify how the Kadison--Schwarz property can hold under
assumptions substantially weaker than complete positivity and yield a structural criterion for the KS property on $M_3$ in terms of Bloch parameters.
\end{abstract}

\maketitle

\section{Introduction}
\label{sec:intro}

Positive linear maps between matrix algebras play a fundamental role
in quantum information theory, operator algebras, and the theory of
open quantum systems.
Among various classes of positive maps, completely positive (CP) maps
occupy a distinguished position due to their physical interpretation
as quantum channels \cite{Choi1975,Kraus1971,Stinespring1955}.
However, in several contexts---notably in the analysis of quantum
dynamical maps, divisibility, and non-Markovian dynamics---one encounters
broader classes of positive maps for which complete positivity is not
required, see, e.g., Refs.~\cite{Breuer2009,Rivas2010,Wolf2008,Breuer2016RMP,Chruscinski2018PRL}.

An important example of such a class is formed by Kadison--Schwarz (KS)
maps.
These maps are characterized by a quadratic operator inequality and
occupy an intermediate position between positivity and complete
positivity.
While every completely positive map satisfies the Kadison--Schwarz
property, the converse implication is false.
As a result, KS maps constitute a genuinely intermediate class of
positive maps with properties that are neither purely algebraic nor
fully captured by complete positivity.

Several general features of KS maps are known.
In particular, unital positive maps satisfy the Kadison--Schwarz
inequality when restricted to normal operators, whereas explicit
counterexamples show that unitality and positivity alone do not
guarantee the KS property.
Beyond such general observations, however, explicit analytic sufficient conditions ensuring the KS property remain scarce and are typically restricted
to low-dimensional or highly symmetric settings.

In this work we study the Kadison--Schwarz property within the Bloch--Gell--Mann
representation of unital maps on $M_d$.
This parametrization is adapted to the $\mathfrak{su}(d)$ structure and makes it
possible to exploit unitary equivalence and canonical forms of the Bloch matrix.
While the representation exists in any dimension, obtaining explicit analytic
KS criteria is strongly dimension dependent.

We focus on $M_3$ and derive an explicit \emph{sufficient} condition for the KS property
for unital positive maps with diagonal Bloch matrix.
The proof is based on an explicit expansion of the KS expression in the Gell--Mann basis
and on a dominance estimate in which the traceless part is controlled by the spectral spread
$\max_{i,j}|\mu_i-\mu_j|$ of the Bloch parameters.
A key structural mechanism is that for diagonal Bloch matrices all contributions involving the
antisymmetric structure constants cancel, so that the relevant estimates are governed by the
symmetric tensor $d_{ijk}$.
We also discuss an illustrative two-parameter diagonal family, which makes the resulting KS region
in parameter space transparent and highlights the difference between positivity, the KS property,
and complete positivity.

The paper is organized as follows.
In Sec.~\ref{sec:ks} we recall the definition of Kadison--Schwarz maps
and review their basic properties.
In Sec.~\ref{sec:gellmann} we introduce the Bloch--Gell--Mann
representation of unital maps on $M_d$.
Our main results, specialized to the case of $M_3$, are presented in
Sec.~\ref{sec:main}, followed by an illustrative example.
Technical derivations are collected in the Appendix.

\section{Kadison--Schwarz maps}
\label{sec:ks}

Let $\Phi : M_d \to M_d$ be a linear map.
We say that $\Phi$ is \emph{positive} if $\Phi(X) \ge 0$ for all $X \ge 0$,
and \emph{unital} if $\Phi(\I)=\I$.
The Kadison--Schwarz (KS) property is defined by the operator inequality
\begin{equation}
\Phi(X^\dagger X) \;\ge\; \Phi(X^\dagger)\,\Phi(X),
\label{eq:KS}
\end{equation}
which is required to hold for all $X \in M_d$.
The inequality \eqref{eq:KS} was originally introduced by Kadison
in the context of operator algebras \cite{Kadison1952}.
We emphasize that the KS property does not, in general, imply unitality,
although the two notions are often considered together.

Every completely positive (CP) map satisfies the Kadison--Schwarz inequality,
but the converse implication does not hold.
Hence KS maps form a strictly larger class than CP maps and interpolate
between complete positivity and positivity,
\begin{equation}
\CP \;\subsetneq\; \KS \;\subsetneq\; \Pos,
\end{equation}
see, e.g.,  Refs.~\cite{Chruscinski2010,MukhamedovAbduganiev2010,Mukhamedov2011}.
Despite this simple ordering, the structure of KS maps is considerably
less understood than that of completely positive maps.

We next recall a basic property of unital positive maps that will be
useful in the analysis of the Kadison--Schwarz condition.

\begin{lemma}[Kadison-type contraction]
\label{lem:kadison}
Let $\Phi : M_d \to M_d$ be a unital positive map.
Then, for every Hermitian operator $A \in M_d$,
\begin{equation}
\|\Phi(A)\| \;\le\; \|A\|,
\end{equation}
where $\|\cdot\|$ denotes the operator norm.
\end{lemma}

This estimate follows from Kadison's inequality
\begin{equation}
\Phi(A)^2 \le \Phi(A^2)
\qquad (A=A^\dagger),
\end{equation}
which holds for unital positive maps
\cite{Kadison1952,Paulsen2002,Takesaki2002}.
Indeed, positivity implies that $\Phi$ maps Hermitian elements to Hermitian elements, hence $\Phi(A)=\Phi(A)^\dagger$.
Therefore $\|\Phi(A)\|^2=\|\Phi(A)^2\|$, and Kadison's inequality gives
\[
\|\Phi(A)\|^2=\|\Phi(A)^2\|\le \|\Phi(A^2)\|.
\]
Since $A^2\ge 0$ and $A^2\le \|A^2\|\I$, positivity and unitality imply
$0\le \Phi(A^2)\le \|A^2\|\Phi(\I)=\|A^2\|\I$, hence
$\|\Phi(A^2)\|\le \|A^2\|=\|A\|^2$.

Lemma~\ref{lem:kadison} implies that unital positive maps are
contractions on the self-adjoint part of $M_d$ with respect to the
operator norm.
Importantly, this property does not rely on complete positivity
and plays a key role in the analysis of KS maps
\cite{Chruscinski2010}.

It is worth noting that for unital positive maps the Kadison--Schwarz
inequality is automatically satisfied when restricted to normal
operators, whereas possible violations of the KS condition necessarily
involve non-normal elements
\cite{Chruscinski2010,Mukhamedov2011}.

We now recall a standard example illustrating that unital positive maps
need not satisfy the Kadison--Schwarz inequality.

The transposition map $T : M_d \to M_d$, defined by $T(X)=X^T$, is positive
and unital, but it does not satisfy the Kadison--Schwarz property.
Indeed, for a rank-one operator $X = |1\rangle\langle 2|$ one has
\[
X^\dagger X = |2\rangle\langle 2|, \qquad
X X^\dagger = |1\rangle\langle 1|,
\]
which leads to
\[
T(X^\dagger X) = |2\rangle\langle 2|, \qquad
T(X^\dagger)\,T(X) = |1\rangle\langle 1|.
\]
Hence
\[
T(X^\dagger X) - T(X^\dagger)T(X)
=
|2\rangle\langle 2| - |1\rangle\langle 1|.
\]
The latter operator has eigenvalues $+1$ and $-1$, and therefore is not
positive. In particular, the presence of a negative eigenvalue shows that
\[
T(X^\dagger X) \ngeq T(X^\dagger)T(X),
\]
so the Kadison--Schwarz inequality is violated.
This example is well known and discussed in detail in the literature,
see, e.g., Refs.~\cite{Chruscinski2010,Mukhamedov2011}.

The transposition map therefore provides a simple illustration that the
Kadison--Schwarz condition is genuinely stronger than positivity, even
in the unital case, and motivates the search for explicit criteria
distinguishing KS maps from merely positive ones.

\section{Bloch--Gell--Mann representation of unital maps on $M_d$}
\label{sec:gellmann}

The Bloch representation of linear maps does not depend on the choice of an orthonormal basis of
$\mathfrak{su}(d)$: different choices (e.g., Weyl or generalized Pauli bases) are related by orthogonal
transformations in Bloch space and yield equivalent descriptions of unital maps
\cite{Kimura2003,BengtssonZyczkowski}.
In particular, the Kadison--Schwarz property is invariant under such basis changes.
We use the Gell--Mann basis because its generators are Hermitian and its symmetric and antisymmetric
structure constants allow a transparent treatment of products appearing in the KS inequality
\cite{Georgi1999}.

\subsection{Bloch representation on $M_d$}

Let $\{\lambda_k\}_{k=1}^{d^2-1}$ be a fixed orthonormal basis of
$\mathfrak{su}(d)$ consisting of Hermitian, traceless matrices, normalized as
\begin{equation}
\Tr(\lambda_i \lambda_j) = 2\,\delta_{ij}.
\end{equation}
The symmetric and antisymmetric structure constants $d_{ijk}$ and $f_{ijk}$
are defined in the standard way via the anticommutation and commutation
relations of the generators, respectively.
Together with $\lambda_0 := \sqrt{\tfrac{2}{d}}\,\I$, this basis provides an
orthonormal basis of $M_d$ with respect to the Hilbert--Schmidt inner product.

Any operator $X \in M_d$ can be written in the Bloch form
\begin{equation}
X = w_0 \lambda_0 + \sum_{k=1}^{d^2-1} w_k \lambda_k,
\label{eq:blochX}
\end{equation}
where $w_0 \in \R$ and $w_k \in \C$.
For Hermitian operators $X$, all coefficients $w_k$ are real
\cite{Kimura2003,BengtssonZyczkowski}.

The product of two basis elements can be expressed as
\begin{equation}
\lambda_i \lambda_j
=
\frac{2}{d}\,\delta_{ij}\,\I
+ \sum_{k=1}^{d^2-1}
\left(
d_{ijk} + i f_{ijk}
\right)\lambda_k,
\label{eq:structure}
\end{equation}
where $f_{ijk}$ and $d_{ijk}$ are the completely antisymmetric and symmetric
structure constants of $\mathfrak{su}(d)$, respectively
\cite{Georgi1999,Slansky}.

\subsection{Unital maps in the Bloch representation}

Let $\Phi : M_d \to M_d$ be a unital linear map.
In the Bloch representation \eqref{eq:blochX}, such a map acts affinely as
\begin{equation}
\Phi(X)
=
w_0 \lambda_0 + \sum_{k=1}^{d^2-1} (T w)_k \lambda_k,
\label{eq:blochmap}
\end{equation}
where $w=(w_1,\dots,w_{d^2-1})^T$ and $T$ is a real $(d^2-1)\times(d^2-1)$
matrix, referred to as the Bloch matrix of $\Phi$
\cite{Kimura2003,Ruskai2002}.
Unitality of $\Phi$ implies that the identity component $w_0$ is preserved,
while positivity imposes nontrivial constraints on $T$.

The Bloch matrix representation is not unique.
If $\Phi$ is conjugated by unitary operators,
\begin{equation}
\Phi \;\mapsto\; \mathcal{U}_1 \circ \Phi \circ \mathcal{U}_2,
\qquad
\mathcal{U}_i(X) = U_i X U_i^\dagger,
\end{equation}
then the corresponding Bloch matrix transforms as
\begin{equation}
T \;\mapsto\; O_1\, T\, O_2,
\end{equation}
where $O_1,O_2 \in SO(d^2-1)$ are orthogonal matrices induced by the adjoint 
action of $SU(d)$ on $\mathfrak{su}(d)$
\cite{BengtssonZyczkowski}.

\subsection{Canonical form}

Using the polar decomposition, any Bloch matrix $T$ can be written as
\begin{equation}
T = R S,
\end{equation}
where $R \in O(d^2-1)$ is orthogonal and $S$ is symmetric and positive
semidefinite.
If the orthogonal factor $R$ is induced by unitary pre- and post-processing,
i.e., by conjugations $X\mapsto UXU^\dagger$ acting on the traceless sector,
then the Kadison--Schwarz property is preserved under the corresponding
transformation $T \mapsto O_1 T O_2$.
For $d\ge 3$, however, not every orthogonal transformation in $SO(d^2-1)$ is realized
by the adjoint action of $SU(d)$.
In what follows we therefore focus on the analytically tractable class of
unital maps that are unitarily equivalent (within this physical equivalence)
to maps with symmetric, and in particular diagonal, Bloch matrices.

Accordingly, we focus on unital maps whose Bloch matrices can be chosen symmetric within their unitary equivalence class.
In particular, if such a symmetric Bloch matrix is diagonalizable, then the map
$\Phi$ is unitarily equivalent to a map with diagonal Bloch matrix
\begin{equation}
T = \diag(\mu_1,\mu_2,\dots,\mu_{d^2-1}),
\label{eq:diagonalT}
\end{equation}
where the real parameters $\mu_k$ characterize the action of $\Phi$ on the
traceless subspace.

This canonical form provides a convenient starting point for analyzing
positivity and Kadison--Schwarz conditions.
In the following section we specialize this framework to the case of
$M_3$, where the explicit structure of $\mathfrak{su}(3)$ allows for
analytic estimates.

\section{Main result: Kadison--Schwarz condition on $M_3$}
\label{sec:main}

In this section we specialize the general Bloch--Gell--Mann framework
developed above to the case of the matrix algebra $M_3$.
The special role of $M_3$ stems from the explicit structure of the
Lie algebra $\mathfrak{su}(3)$, which allows one to control the
structure constants appearing in operator products and to derive
analytic estimates for the Kadison--Schwarz inequality.

\begin{theorem}
\label{thm:KS_M3}
Let $\Phi : M_3 \to M_3$ be a unital positive linear map with diagonal
Bloch matrix
\[
T = \diag(\mu_1,\mu_2,\dots,\mu_8)
\]
in the Gell--Mann representation.
Then, by Lemma~\ref{lem:kadison}, one has
\begin{equation}
\max_{1\le k \le 8} |\mu_k| \le 1,
\label{eq:contraction}
\end{equation}
since $\lambda_k=\lambda_k^\dagger$ and $\Phi(\lambda_k)=\mu_k\lambda_k$, hence
\[
|\mu_k|\,\|\lambda_k\|=\|\Phi(\lambda_k)\|\le \|\lambda_k\|
\]
and therefore $|\mu_k|\le 1$ since $\|\lambda_k\|>0$.
Define
\begin{equation}
c_3(\mu)
:=
\inf_{X\neq 0}\frac{\alpha(X)}{\|X^\dagger X\|}\ \ge 0,
\label{eq:c3_def}
\end{equation}
where $\alpha(X)$ is the scalar coefficient in the expansion
\eqref{eq:KS_expansion}.
Note that $\|X^\dagger X\|>0$ for $X\neq 0$, so $c_3(\mu)$ is well-defined.
If
\begin{equation}
\max_{i,j} |\mu_i - \mu_j|
\;\le\;
\frac{c_3(\mu)}{C_3},
\label{eq:KS_condition}
\end{equation}
where $C_3$ is the structural constant from Appendix~\ref{app:C3}, then
$\Phi$ satisfies the Kadison--Schwarz inequality.
\end{theorem}
\begin{remark}
The constant $C_3$ controls the traceless contribution in the expansion
\eqref{eq:KS_expansion} via the norm estimate derived in Appendix~\ref{app:C3}.
The quantity $c_3(\mu)$ is defined in \eqref{eq:c3_def} and quantifies a
uniform lower bound for the scalar term $\alpha(X)$ relative to $\|X^\dagger X\|$.
Condition~\eqref{eq:KS_condition} is sufficient, but in general not necessary,
for the Kadison--Schwarz property within the class of unital positive maps.
Note that if $c_3(\mu)=0$, then \eqref{eq:KS_condition} forces $\max_{i,j}|\mu_i-\mu_j|=0$, so the sufficient criterion becomes nontrivial only when $c_3(\mu)>0$.
Although $c_3(\mu)$ is defined variationally, in concrete subclasses of diagonal maps it can often be bounded from below explicitly; we leave optimization of such bounds for future work.
\end{remark}
We stress that the theorem provides a purely analytic sufficient criterion that does not rely on complete positivity or optimization-based arguments, and therefore isolates a genuinely structural mechanism behind the Kadison--Schwarz property on $M_3$.
\subsection{Reduction of the Kadison--Schwarz inequality}

Let $X \in M_3$ be an arbitrary operator and write it in the
Bloch--Gell--Mann form
\[
X = w_0 \lambda_0 + \sum_{k=1}^8 w_k \lambda_k.
\]
Using the product relations for the Gell--Mann generators, the operator
$X^\dagger X$ can be expressed as a linear combination of the identity
and traceless generators.
Applying the unital map $\Phi$ with diagonal Bloch matrix
$T=\diag(\mu_1,\dots,\mu_8)$, the Kadison--Schwarz expression takes the form
\begin{equation}
\Phi(X^\dagger X) - \Phi(X^\dagger)\Phi(X)
=
\alpha(X)\,\I + \sum_{k=1}^8 \beta_k(X)\,\lambda_k,
\label{eq:KS_expansion}
\end{equation}
where the coefficients $\alpha(X)$ and $\beta_k(X)$ depend quadratically
on the Bloch components $w_k$ and on the parameters $\mu_k$.

\begin{lemma}[Cancellation of the antisymmetric part for diagonal $T$]
\label{lem:f_cancel}
Let $\Phi$ have diagonal Bloch matrix $T=\diag(\mu_1,\dots,\mu_8)$ with real parameters $\mu_k$.
Then in the expansion of the Kadison--Schwarz expression
$\Phi(X^\dagger X)-\Phi(X^\dagger)\Phi(X)$
all contributions proportional to the antisymmetric structure constants $f_{ijk}$ cancel.
\end{lemma}

\begin{proof}
In the Bloch--Gell--Mann product expansion, the $f_{ijk}$-terms appear with coefficients antisymmetric
under the exchange $i\leftrightarrow j$, whereas the quadratic combinations $w_i\overline{w_j}$ arising
from $X^\dagger X$ are paired with their conjugates $w_j\overline{w_i}$. This pairing is automatic because the KS expression is Hermitian and the Bloch expansion collects each coefficient together with its complex conjugate counterpart.
Since $T$ is diagonal with real $\mu_k$, the resulting coefficients are symmetric in $(i,j)$, and hence
the total antisymmetric contribution vanishes after summation.
\end{proof}

By Lemma~\ref{lem:f_cancel}, for diagonal $T$ the KS expression is governed entirely by the symmetric tensor $d_{ijk}$.
(For $d=2$ one has $d_{ijk}\equiv 0$, so this mechanism is absent in the qubit case.)

\subsection{Proof of Theorem~\ref{thm:KS_M3}}

In view of the reduction described above, the Kadison--Schwarz condition reduces to the positivity of the operator \eqref{eq:KS_expansion}.
We estimate separately its scalar and traceless parts.

\medskip
\noindent
\textit{Scalar contribution.}
Using the product relations for the Gell--Mann generators one verifies that
the coefficient $\alpha(X)$ is a sum of terms of the form
\[
(1-\mu_k^2)\,|w_k|^2
\]
(together with nonnegative contributions proportional to $|w_0|^2$).
Indeed, the identity component in \eqref{eq:structure} is proportional to $\delta_{ij}$, hence contributes only terms with $i=j$.
Hence, under the contraction assumption \eqref{eq:contraction},
one has $\alpha(X)\ge 0$ for all $X$.

Moreover, define
\[
c_3(\mu):=\inf_{X\neq 0}\frac{\alpha(X)}{\|X^\dagger X\|}\ge 0,
\]
as in \eqref{eq:c3_def}. By definition, this implies the uniform bound
\begin{equation}
\alpha(X)\ \ge\ c_3(\mu)\,\|X^\dagger X\|
\qquad\text{for all }X\in M_3.
\label{eq:alpha_lower_proof}
\end{equation}

\medskip
\noindent
\textit{Traceless contribution.}
The remaining traceless part can be estimated using operator norm bounds
for expressions involving the symmetric structure constants $d_{ijk}$.
More precisely, one obtains
\begin{equation}
\biggl\|
\sum_{k=1}^8 \beta_k(X)\,\lambda_k
\biggr\|
\;\le\;
C_3 \, \max_{i,j} |\mu_i - \mu_j| \, \|X^\dagger X\|,
\label{eq:beta_bound}
\end{equation}
where $C_3$ depends only on the algebraic structure of $\mathfrak{su}(3)$,
as shown in Appendix~\ref{app:C3}.

\medskip
\noindent
\textit{Dominance argument.}
If
\[
\max_{i,j} |\mu_i-\mu_j| \;\le\; \frac{c_3(\mu)}{C_3},
\]
then combining \eqref{eq:alpha_lower_proof} and \eqref{eq:beta_bound} yields
\[
\biggl\|
\sum_{k=1}^8 \beta_k(X)\,\lambda_k
\biggr\|
\;\le\;
c_3(\mu) \, \|X^\dagger X\|
\;\le\;
\alpha(X).
\]
Therefore the traceless part is dominated by the positive scalar part,
and the operator \eqref{eq:KS_expansion} is positive.
Hence,
\[
\Phi(X^\dagger X) \ge \Phi(X^\dagger)\Phi(X)
\]
for all $X \in M_3$, which proves the theorem.
A detailed derivation of the bound \eqref{eq:beta_bound} and an explicit
expression for the constant $C_3$ are provided in the Appendix.
From a structural perspective, the Kadison--Schwarz condition for diagonal Bloch maps on $M_3$ can be interpreted as a stability property under controlled spectral anisotropy.
\section{Illustrative diagonal family}
\label{sec:example}

To illustrate the content of Theorem~\ref{thm:KS_M3},
consider a two-parameter family of diagonal Bloch matrices
\begin{equation}
T(t,s)=\diag(t,t,t,t,t,t,t,s),
\end{equation}
with real parameters $t,s \in [-1,1]$. 
For general $(t,s)$ the associated unital map need not be positive.
We do not characterize the positivity region of $T(t,s)$ here and restrict the discussion to parameters for which the map is positive.
Such maps act isotropically on a seven-dimensional
subspace of $\mathfrak{su}(3)$ while allowing an
independent deformation along a distinguished direction.

The contraction condition \eqref{eq:contraction}
is satisfied whenever $|t|\le1$ and $|s|\le1$.
Moreover,
\begin{equation}
\max_{i,j}|\mu_i-\mu_j| = |t-s|.
\end{equation}
Hence Theorem~\ref{thm:KS_M3} guarantees the
Kadison--Schwarz property whenever
\begin{equation}
|t-s| \le \frac{c_3(t,s)}{C_3},
\end{equation}
where $C_3$ is the structural constant defined in
Appendix~\ref{app:C3} and
\[
c_3(t,s):=c_3(\mu)\ \text{with $c_3(\mu)$ given by \eqref{eq:c3_def}, for}\quad \mu=(t,t,t,t,t,t,t,s).
\]

This example shows explicitly that the sufficient
condition defines a nontrivial region in parameter space,
controlled only by the relative spectral spread of the
Bloch parameters rather than by their individual values.
Geometrically, the admissible region corresponds to a
neighborhood of the isotropic line $t=s$ whose width is
determined by the ratio $c_3(t,s)/C_3$, reflecting the balance
between the positive scalar part and the traceless
fluctuations governed by the $\mathfrak{su}(3)$ structure
constants.

\medskip
\noindent
\textit{Positivity and complete positivity for the isotropic subfamily.}
For the isotropic choice $t=s$, the map acts as a scalar on the
traceless subspace and admits the depolarizing form
\begin{equation}
\Phi_t(X)= tX + (1-t)\frac{\Tr(X)}{3}\,\I.
\end{equation}

Applying $\Phi_t$ to a rank-one projection
$P_\psi=|\psi\rangle\langle\psi|$ yields
\[
\Phi_t(P_\psi)= tP_\psi + \frac{1-t}{3}\I,
\]
whose eigenvalues are
$\lambda_\parallel=(2t+1)/3$ and
$\lambda_\perp=(1-t)/3$ (with multiplicity $2$).
Hence the map is positive if and only if 
\begin{equation}
-\frac12 \le t \le 1,
\end{equation}
see \cite{Watrous2018,NielsenChuang}.
The corresponding Choi matrix
$J(\Phi_t)=(\Phi_t\otimes\id)(|\Phi\rangle\langle\Phi|)$,
with $|\Phi\rangle=\frac1{\sqrt{3}}\sum_{i=1}^3|ii\rangle$,
takes the form
\begin{equation}
J(\Phi_t)
=
t\,|\Phi\rangle\langle\Phi|
+
\frac{1-t}{9}\,\I\otimes\I.
\end{equation}
Its spectrum consists of the eigenvalue
\[
\lambda_1=\frac{8t+1}{9}
\]
and the eigenvalue
\[
\lambda_2=\frac{1-t}{9}
\]
with multiplicity $8$.
Therefore $\Phi_t$ is completely positive if and only if 
\begin{equation}
-\frac18 \le t \le 1, 
\end{equation}
see \cite{Watrous2018,NielsenChuang,Holevo2001}.
\begin{remark}
Theorem~\ref{thm:KS_M3} is formulated for unital positive maps.
Accordingly, in the present example the Kadison--Schwarz condition
is interpreted only within the parameter region where the map
$\Phi_t$ is positive, namely $-\tfrac12 \le t \le 1$.
Although the spectral condition of the theorem is trivially satisfied
for all $t$ in the isotropic case ($t=s$), values $t<-1/2$
are excluded since positivity is a necessary prerequisite for the
Kadison--Schwarz property.
\end{remark}

For the same isotropic subfamily one has
$\max_{i,j}|\mu_i-\mu_j|=0$, and hence
Theorem~\ref{thm:KS_M3} guarantees the
Kadison--Schwarz property throughout the region where the
map is positive, i.e.\ for $-\tfrac12 \le t \le 1$.
Consequently, within the class of positive unital maps,
the sufficient KS region strictly extends beyond the
completely positive regime.

The family $T(t,s)$ therefore provides a geometric
illustration of how analytic bounds derived from the
$\mathfrak{su}(3)$ structure control deviations from
complete positivity while still ensuring the
Kadison--Schwarz property.
\section{Conclusions}
We have derived an explicit sufficient criterion for the Kadison--Schwarz (KS) property of unital positive maps on $M_3$ within the Bloch--Gell--Mann representation.
For diagonal Bloch matrices $T=\diag(\mu_1,\dots,\mu_8)$ satisfying the contraction bound
$\max_k|\mu_k|\le 1$, the KS condition follows whenever the spectral spread
$\max_{i,j}|\mu_i-\mu_j|$ is small enough compared to the ratio $c_3(\mu)/C_3$,
where $c_3(\mu)$ quantifies the uniform dominance of the scalar term in the KS expansion
and $C_3$ is a structural constant determined solely by the symmetric tensor $d_{ijk}$
of $\mathfrak{su}(3)$.

A key structural feature behind the proof is that for diagonal $T$ all contributions
involving the antisymmetric structure constants $f_{ijk}$ cancel in the KS expression,
reducing the problem to norm estimates governed entirely by $d_{ijk}$.
This yields a transparent dominance argument: the traceless fluctuation term is controlled
by $\max_{i,j}|\mu_i-\mu_j|$ and is dominated by the positive scalar part whenever
\eqref{eq:KS_condition} holds.
In particular, within the positive unital regime the sufficient KS region may extend
strictly beyond complete positivity, illustrating that the KS property captures a
genuinely intermediate level of positivity.

The present analysis suggests several directions for further work.
First, it would be desirable to obtain sharper (and possibly more explicit) lower bounds
on $c_3(\mu)$ for natural subclasses of diagonal maps, leading to more quantitative KS
regions in parameter space.
Second, extending the approach to $M_d$ with $d>3$ requires dimension-dependent control
of expressions involving the symmetric structure constants of $\mathfrak{su}(d)$ and a
careful treatment of unitary equivalence, since not every orthogonal transformation in
$SO(d^2-1)$ arises from the adjoint action of $SU(d)$ (see \cite{BengtssonZyczkowski,Georgi1999}).
We expect that the analytic mechanism used here may be useful in the study of KS
properties of dynamical maps and generators in open quantum systems, as well as in the
systematic exploration of intermediate positivity classes beyond complete positivity.
It would be interesting to understand whether analogous dominance mechanisms persist for broader classes of positive maps beyond the diagonal Bloch setting.
\section*{Acknowledgements}

The author would like to thank Aristides Katavolos for helpful comments
that improved the clarity of the manuscript.
\appendix

\section{Estimates involving the symmetric structure constants of $\mathfrak{su}(3)$}
\label{app:C3}

In this Appendix we provide the technical estimates underlying
Theorem~\ref{thm:KS_M3}.
In particular, we derive bounds on expressions involving the symmetric
structure constants $d_{ijk}$ of $\mathfrak{su}(3)$ and introduce the
constant $C_3$ appearing in Eqs.~\eqref{eq:KS_condition}
and~\eqref{eq:beta_bound}.

\subsection{Structure constants of $\mathfrak{su}(3)$}

Let $\{\lambda_k\}_{k=1}^8$ denote the standard Gell--Mann generators,
normalized as $\Tr(\lambda_i \lambda_j)=2\delta_{ij}$.
Their products satisfy
\begin{equation}
\lambda_i \lambda_j
=
\frac{2}{3}\,\delta_{ij}\,\I
+
\sum_{k=1}^8
\left(
d_{ijk} + i f_{ijk}
\right)\lambda_k,
\end{equation}
where $f_{ijk}$ and $d_{ijk}$ are the antisymmetric and symmetric
structure constants of $\mathfrak{su}(3)$, respectively.
Explicit values of these constants can be found, e.g., in
Refs.~\cite{Georgi1999,Slansky}.

In the analysis of the Kadison--Schwarz inequality only the symmetric
constants $d_{ijk}$ contribute.
This follows from the cancellation of all terms involving $f_{ijk}$ for
diagonal Bloch matrices, as discussed in Sec.~\ref{sec:main}.

\subsection{Norm estimates}

Consider expressions of the form
\begin{equation}
Y
=
\sum_{k=1}^8 y_k \lambda_k,
\end{equation}
where the coefficients $y_k$ arise as quadratic combinations of the
Bloch components $w_i$ and the parameters $\mu_j$.
In particular, when $Y$ is the traceless part of the KS expansion \eqref{eq:KS_expansion},
we have $y_k=\beta_k(X)$ for $k=1,\dots,8$.
Using the submultiplicativity of the operator norm and the orthonormality
of the Gell--Mann basis, one obtains the bound
\begin{equation}
\|Y\|
\;\le\;
\sum_{k=1}^8 |y_k|\,\|\lambda_k\|
\le
\Bigl(\max_{1\le k\le 8}\|\lambda_k\|\Bigr)\sum_{k=1}^8 |y_k|.
\label{eq:norm_basic}
\end{equation}
Since $\max_k\|\lambda_k\|$ is a fixed numerical constant for the chosen normalization,
we incorporate it into the definition of $C_3$ below.
The coefficients $y_k$ contain terms of the form
\begin{equation}
\sum_{i,j=1}^8 (\mu_i-\mu_j)\, d_{ijk}\, w_i \overline{w_j}.
\label{eq:yk_structure}
\end{equation}

To make the dependence explicit, we introduce the auxiliary constant
\begin{equation}
C_3^{(0)}
:=
\sup_{\|u\|_2=\|v\|_2=1}
\sum_{k=1}^8
\left|
\sum_{i,j=1}^8 d_{ijk}\,u_i \overline{v_j}
\right|,
\label{eq:C3_explicit}
\end{equation}
where $\|\cdot\|_2$ denotes the Euclidean norm on $\C^8$. Here $d_{ijk}\in\R$ for the Gell--Mann normalization.
Since the set of nonvanishing symmetric structure constants $d_{ijk}$
is finite, the quantity~\eqref{eq:C3_explicit} is finite and depends
only on the algebraic structure of $\mathfrak{su}(3)$ and the chosen
normalization of the generators.

Applying the Cauchy--Schwarz inequality to~\eqref{eq:yk_structure}
and using~\eqref{eq:C3_explicit}, one obtains
\begin{equation}
\sum_{k=1}^8 |y_k|
\;\le\;
C_3^{(0)} \,
\max_{i,j} |\mu_i-\mu_j| \,
\|w\|_2^2,
\label{eq:w_bound_step}
\end{equation}
where $C_3^{(0)}$ depends only on the symmetric structure constants
$d_{ijk}$ and the chosen normalization of the generators.

To relate $\|w\|_2$ to an operator norm quantity, note that the
Bloch--Gell--Mann expansion
$X = w_0\lambda_0 + \sum_{k=1}^8 w_k\lambda_k$
satisfies, by $\Tr(\lambda_a^\dagger\lambda_b)=2\delta_{ab}$,
\begin{equation}
\|X\|_{\mathrm{HS}}^2
=
2\bigl(|w_0|^2 + \|w\|_2^2\bigr),
\qquad\text{hence}\qquad
\|w\|_2^2 \le \frac12\,\|X\|_{\mathrm{HS}}^2 .
\end{equation}
Moreover, for $X\in M_3$ one has $\|X\|_{\mathrm{HS}} \le \sqrt{3}\,\|X\|$, and therefore
\begin{equation}
\|w\|_2^2 \le \frac12\,\|X\|_{\mathrm{HS}}^2 \le \frac32\,\|X\|^2
= \frac32\,\|X^\dagger X\|.
\label{eq:w_to_opnorm}
\end{equation}
Combining \eqref{eq:w_bound_step} and \eqref{eq:w_to_opnorm} yields
\begin{equation}
\sum_{k=1}^8 |y_k|
\;\le\;
\frac32\,C_3^{(0)} \, \max_{i,j} |\mu_i-\mu_j| \, \|X^\dagger X\|.
\end{equation}

Together with~\eqref{eq:norm_basic} this yields
\begin{equation}
\biggl\|
\sum_{k=1}^8 y_k \lambda_k
\biggr\|
\;\le\;
\Bigl(\max_{1\le k\le 8}\|\lambda_k\|\Bigr)\,
\frac32\,C_3^{(0)}\,
\max_{i,j} |\mu_i-\mu_j| \, \|X^\dagger X\|.
\end{equation}
For notational simplicity we redefine
\begin{equation}
C_3 := \Bigl(\max_{1\le k\le 8}\|\lambda_k\|\Bigr)\,\frac32\,C_3^{(0)},
\end{equation}
which is still a structural constant depending only on $\mathfrak{su}(3)$
(and the chosen normalization), and we arrive at the bound stated in the main text:
\begin{equation}
\biggl\|
\sum_{k=1}^8 y_k \lambda_k
\biggr\|
\;\le\;
C_3 \, \max_{i,j} |\mu_i-\mu_j| \, \|X^\dagger X\|.
\label{eq:C3_def}
\end{equation}

\subsection{Remarks}

The auxiliary constant $C_3^{(0)}$ is the operator norm of the bilinear
form induced by the symmetric tensor $d_{ijk}$ on $\C^8$.
Its finiteness follows from the finite number of nonvanishing
structure constants of $\mathfrak{su}(3)$.
While its precise numerical value is not needed for the formulation
of Theorem~\ref{thm:KS_M3}, the definition~\eqref{eq:C3_explicit}
shows that $C_3^{(0)}$ (and hence also $C_3$ defined above)
depends only on the algebraic data of $\mathfrak{su}(3)$ and is independent of the particular map $\Phi$.
This ensures that the condition~\eqref{eq:KS_condition}
is purely structural.
In contrast, the quantity $c_3(\mu)$ from \eqref{eq:c3_def} depends on the
Bloch parameters and captures how strongly the scalar part dominates in the
Kadison--Schwarz expansion for the given diagonal map.

\bibliography{ks}

\end{document}